\documentclass[11pt]{article}
\usepackage{graphicx}
\usepackage{amsmath,amsfonts,amssymb}
\usepackage{color}
\usepackage{bbold}
\allowdisplaybreaks
\usepackage{charter}
\def\sloppy{\tolerance=100000\hfuzz=\maxdimen\vfuzz=\maxdimen}
\allowdisplaybreaks
\def \beq  {\begin{equation}}
\def \eeq  {\end{equation}}
\def \beqar {\begin{eqnarray}}
\def \eeqar {\end{eqnarray}}
\def\sqr#1#2{{\vcenter{\vbox{\hrule height.#2pt
\hbox{\vrule width.#2pt height#1pt \kern#1pt
\vrule width.#2pt}\hrule height.#2pt}}}}

\def\vx {{\vec x}}

\def\vf {{\varphi}}

\def\Tr {{\rm Tr}}

\def\vx {{\vec x}}

\def\vf {{\varphi}}

\def\del {\partial}

\def\e {\epsilon}

\def\D {{\cal D}}

\def\A {{\cal A}}

\def\half{\textstyle{1\over 2}}

\begin{document}
\def \CMP {{Commun. Math. Phys.}}
\def \PRL {{Phys. Rev. Lett.}}
\def \PL {{Phys. Lett.}}
\def \NPBProc {{Nucl. Phys. B (Proc. Suppl.)}}
\def \NP {{Nucl. Phys.}}
\def \RMP {{Rev. Mod. Phys.}}
\def \JGP {{J. Geom. Phys.}}
\def \CQG {{Class. Quant. Grav.}}
\def \MPL {{Mod. Phys. Lett.}}
\def \IJMP {{ Int. J. Mod. Phys.}}
\def \JHEP {{JHEP}}
\def \PR {{Phys. Rev.}}
\def \JMP {{J. Math. Phys.}}
\def \GRG{{Gen. Rel. Grav.}}
\begin{titlepage}
\null\vspace{-62pt} \pagestyle{empty}
\begin{center}
\rightline{CCNY-HEP-17/4}
\rightline{June 2017}
\vspace{1truein} {\Large\bfseries
Actions for particles and strings and}\\
\vspace{6pt}
\vskip .1in
{\Large \bfseries  Chern-Simons gravity}\\
\vskip .2in
{\Large\bfseries ~}\\
{\large\sc Lei Jiusi} and
 {\large\sc V.P. Nair}\\
\vskip .2in
\vskip .1in
{\itshape Physics Department\\
City College of the CUNY\\
New York, NY 10031}\\
\vskip .1in
\begin{tabular}{r l}
E-mail:&{\fontfamily{cmtt}\fontsize{11pt}{15pt}\selectfont leijiusi@gmail.com}\\
&{\fontfamily{cmtt}\fontsize{11pt}{15pt}\selectfont vpnair@ccny.cuny.edu}
\end{tabular}

\vspace{.8in}
\centerline{\large\bf Abstract}
\end{center}
We consider actions for particles and strings, including twistorial descriptions
on 4d Minkowski and AdS$_5$ spacetimes from the point of view of co-adjoint orbits
for the isometry group. We also consider the collective coordinate dynamics of 
singular solutions in Chern-Simons (CS) theories and CS theories of gravity.
This is a generalization of the work of Einstein, Infeld and Hoffmann
and also has potential points of contact with fluid-gravity correspondence.

\end{titlepage}

\pagestyle{plain} \setcounter{page}{2}

\section{Introduction}
The first part of this article is about exploring interconnections between somewhat different formulations of the actions for point-particles and strings. Point-particles can be identified
as unitary representations of the Poincar\'e group, or, more generally, of the 
appropriate isometry group. It has been rather well known for a long time
that
one can use a co-adjoint orbit action for describing such representations
and, in fact, it is the basic paradigm for the whole idea of 
geometric quantization \cite{geom}. Actions in terms of the spacetime coordinates, in terms of twistor variables, etc., have also been used. More recently, there has been some work on particle dynamics
on AdS spacetimes, including a twistor description for massive particles \cite{ABT}.
We analyze many of these actions from a single point of view, namely, 
in terms of the co-adjoint orbit action for the isometry group.

In the second part of this article, we consider singular vortex or instanton solutions
in Chern-Simons (CS) theories, including the special important case of
CS theories of gravity \cite{zanelli}. Removing the locations of the singularities from the manifold
under consideration, one can obtain a nonsingular description. The CS action 
for such solutions is shown to reduce to an appropriate co-adjoint orbit action,
thus recovering the results of the earlier sections from a different point of view.
Conceptually, this is similar to the work of Einstein, Infeld and Hoffmann (EIH), who considered 
point-particles as singularities of the gravitational field and then showed that
the multiparticle dynamics is determined by the field equations of general
relativity \cite{EIH}. Our approach is similar in spirit, and, in fact, may be considered as
the EIH idea applied to CS gravity.
Our analysis also falls within the circle of ideas related
to recent work on fluid-gravity correspondence \cite{fluid-g}.
Here one considers a general diffeomorphism of special solutions and then,
viewing the diffeomorphism as providing collective degrees of freedom,
one obtains
the evolution equations for these collective modes,
which are seen to be essentially the
equations of fluid dynamics. Since one can define fluid dynamics in terms
of Poincar\'e representations as well \cite{{fluid-us}, {KN}}, we expect that there will be further linkages
of our work to
the fluid-gravity correspondence.

This paper is organized as follows.
In section 2, we will consider the 
actions for massive and massless
particles, the Nambu-Goto string and the null string in 4d Minkowski space
in terms
of the Poincar\'e group. We will relate our results to the
actions which have been suggested for these cases in 
different contexts in the literature.
In section 3, we will do a similar analysis for the AdS$_5$ spacetime.
recovering among other things,
the twistor description of massive particles
obtained in \cite{ABT}. We will also consider the twistor version for massless particles and 
null strings
in AdS$_5$.
In section 4, we will consider the Chern-Simons theory for an arbitrary gauge group
$G$ to show how singular solutions can lead to the co-adjoint orbit action.
A similar analysis for
Chern-Simons gravity in 3 and 5 dimensions will also be discussed.

\section{Poincar\'e orbits}

Free particles and free strings can be described in terms of representations of the
Poincar\'e group. For particles, the  action is given by the integral of the symplectic
potential on a co-adjoint orbit of the spacetime isometry group. 
So we start with a statement of this method.
Let $g$ be a matrix representing a
general element of a Lie group $G$ (in some particular matrix representation).
The symplectic potential on an orbit is given by
 \beq
 \A = i \sum_a w_a \, \Tr ( t_a \, g^{-1} { d g} )
 \label{2}
 \eeq
  where $t_a$ give a basis of the diagonal generators of the Lie algebra (the Cartan subalgebra)
 and $w_a$ are a set of numbers characterizing the chosen orbit.
 We take $t_a$ to be normalized
 as $\Tr (t_a\,t_b ) = \delta_{ab}$.
The action for free particle dynamics can then be taken to be
 \beq
 S = i  \sum_a w_a \, \int d\tau ~ \Tr ( t_a \, g^{-1} {\dot g} ),
 \hskip .3in {\dot g} = {d g \over d \tau}
 \label{1}
 \eeq
 where the integral is along some path  $g (\tau)$ in the group parametrized by $\tau$.
Upon quantization, the theory defined by (\ref{1})
leads to a Hilbert space
 which carries a unitary irreducible representation (UIR) of $G$, this UIR being specified by the highest weight $(w_1, w_2, \cdots, w_r)$. Here $r$ is the rank of the group, which is also the range of summation for
 the subscript $a$. The canonical one-form associated to (\ref{1}) is evidently $\A$.
 Under transformations $g \rightarrow g \, \exp (-i t_a \vf^a )$, we find
 $\A \rightarrow \A + d f $, $f = \sum w_a \vf^a$.
 Thus the symplectic two-form $\Omega = d \A$ is invariant under
 the transformation $g \rightarrow g \, \exp (-i t_a \vf^a )$ and
  hence $\Omega$ is defined on $G/H_{\rm C}$, $H_{\rm C}$ being the 
 Cartan subgroup.
 Further, the transformation $\A \rightarrow \A + df$ shows that in the quantum theory, where wave functions transform as $e^{iS}$, there can be quantization conditions on $w_\alpha$.
These will turn out to be the required
conditions for $(w_1, w_2, \cdots, w_r)$ to qualify as the highest weight
 of a UIR. The existence of such conditions will, of course,
 depend on whether the corresponding directions in $H_{\rm C}$
 are compact or not.

We will now apply this, taking $G$ to be the Poincar\'e group in 4-dimensional Minkowski space
and in AdS$_5$, to obtain point-particle and string actions. We will also relate this
to the twistor actions for particles and strings.

For particle dynamics in 4-dimensional Minkowski space, it is simplest to consider
the Poincar\'e group as a contraction of the de Sitter group. We can use the standard
spinorial representation with the generators
\beq
P_\mu = {\gamma_\mu \over l}, \hskip .3in
J_{\mu\nu} = {i \over 4} [\gamma_\mu, \gamma_\nu]
\label{3}
\eeq
where $\gamma_\mu$ are the $4\times 4$ Dirac matrices and $l$ is a parameter with
the dimension of length. The limit $l \rightarrow \infty$ is the contraction giving 
the Poincar\'e algebra with commuting translation generators $P_\mu$.
For a massive point-particle, we need an orbit that corresponds to a time-like momentum
vector, so we can take this as $t_1 \sim \gamma_0$.There are three
generators $J_{12}$, $J_{23}$, $J_{31}$ which commute with $\gamma_0$.
We can take $h_2$ as
one of these, say, $\gamma_1 \gamma_2$.
The symplectic potential for a massive particle in 4d Minkowski spacetime is thus
\beq
\A =  \biggl[ -i { m l \over 4} \Tr (\gamma_0\, g^{-1} dg ) 
+ {s \over 2} \Tr (\gamma_1 \gamma_2\, g^{-1} dg)\biggr]_{l \rightarrow \infty}
\label{4}
\eeq
A general element of the de Sitter group can be parametrized as
\beq
g = \exp\left( i {\gamma_\mu x^\mu \over l}\right)~ \Lambda
\label{5}
\eeq
where $\Lambda$ denotes an element of the Lorentz group, of the form
$\Lambda = \exp( i J_{\mu\nu} \theta^{\mu\nu})$. Using this parametrization,
the symplectic potential (\ref{4}) reduces to
\beq
\A = p_\mu \, dx^\mu + {s\over 2} \Tr (\gamma_1 \gamma_2 \, \Lambda^{-1} d \Lambda )
\label{6}
\eeq
where $p_\mu = m \Lambda^{\alpha}_{~0} \eta_{\mu\alpha}$ and $\Lambda^{\alpha}_{~\beta}$
is the vector representation of the Lorentz group defined by
$\Lambda \gamma_\beta \Lambda^{-1} = \gamma_\alpha\,  \Lambda^\alpha_{~\beta}$.
Notice that by construction $p^2 = m^2$. $\A$ given in (\ref{6}) is the standard and rather well-known form used for describing point-particles with mass and spin.
The second term in $\A$ describes the spin degrees of freedom.
A variant of this formalism is to consider
$p_\mu$ as four independent variables to begin with, i.e., not given 
in terms of $\Lambda^{\alpha}_{~0}$, and then impose
the condition $p^2 = m^2 $ as a constraint.
Further points about the dynamics, including coupling to external fields, 
the emergence of the wave equation in the quantum theory,
magnetic moment and spin-orbit interactions, extensions to fluids, etc. can be found
in \cite{KN} as well as in earlier references cited there.

For a massless point-particle, we need a null orbit. This can be obtained by the
choice $t_1 \sim \gamma_0 + \gamma_3$ and $t_2 \sim \gamma_1 \gamma_2$.
Thus
\beqar
\A &=&  \biggl[ -i { \mu l \over 4} \Tr \bigl[ (\gamma_0 + \gamma_3 )\, g^{-1} dg \bigr]
+ {s \over 2} \Tr (\gamma_1 \gamma_2\, g^{-1} dg)\biggr]_{l \rightarrow \infty}
\nonumber\\
&=&\mu\, \eta_{\mu\alpha} (\Lambda^{\alpha}_{~0}+ \Lambda^{\alpha}_{~3} )\, dx^\mu
+ {s\over 2} \Tr (\gamma_1 \gamma_2 \, \Lambda^{-1} d \Lambda )
\label{7}
\eeqar
where $\mu$ is a scale parameter with the dimensions of energy. 
Notice that the momentum
$p_\mu = \mu\, \eta_{\mu\alpha} (\Lambda^{\alpha}_{~0} + \Lambda^{\alpha}_{~3} )$
satisfies $p^2 = 0$. 
Infinitesimal Lorentz transformations acting on the left of $\Lambda$ are given by
\beq
\Lambda \rightarrow (1 + {\textstyle{1\over 4} }\omega^{\alpha \beta} [\gamma_\alpha, \gamma_\beta]
) \, \Lambda
\label{8}
\eeq
while the vectors transform as $p_\mu \rightarrow p_\mu + \omega_{\mu \nu} 
p^\nu$. The canonical generator of this transformation given by the
symplectic form in (\ref{7}) is
\beq
M_{\alpha \beta} = - (x_\alpha p_\beta - x_\beta p_\alpha) + {s \over 8}
\Tr ( \gamma_1 \gamma_2 \Lambda^{-1} [\gamma_\alpha, \gamma_\beta] \Lambda )
\label{9}
\eeq
The Pauli-Lyubanski spin vector can now be calculated as
\beq
W^\mu \equiv -{1\over 2} \epsilon^{\mu\nu\alpha\beta} \, M_{\nu \alpha} \, p_\beta
= s\, p^\mu
\label{10}
\eeq
This identifies $s$ as the helicity of the massless particle.

It is also useful to obtain the twistor action for a massless particle from this description.
For simplicity, we will consider the spinless case first; in this case, the
symplectic form is just the first term of $\A$ in (\ref{7}).
We can reduce this further. Since $\gamma_5$ is invariant under Lorentz transformations,
it is possible to consider the projections ${\half} (1\pm i \gamma_5)$ separately.
We then take $P_\mu = {\half} (1- i \gamma_5) \gamma_\mu/l$.
This will give $P^2 =0$; for a massless particle this is acceptable.
We choose a representation of $\gamma$-matrices as
\beq
\gamma_0 = \left( \begin{matrix} 0&1\\ 1&0\\ \end{matrix} \right),
\hskip .2in
\gamma_i = \left( \begin{matrix} 0&-\sigma_i\\ \sigma_i&0\\ \end{matrix} \right),
\hskip .2in
\gamma_5 = - i\,\left( \begin{matrix} 1&0\\ 0&-1\\ \end{matrix} \right)
\label{13}
\eeq
(Our choice of an antihermitian $\gamma_5$ is convenient for later
discussions.)
The parametrization of the group element may be taken as
\beq
g = \left( \begin{matrix} 1&0\\ i\,X &1\\ \end{matrix} \right)\,
\Lambda,\, \hskip .3in X ={x^0+ \sigma\cdot \vx\over l}
\label{13a}
\eeq
The symplectic form is then obtained as
\beq
\A =  - {i\over 2}~ { \mu l}~ \Tr \bigl[ (\gamma_0 + \gamma_3 )\, g^{-1} dg \bigr]
\label{11} 
\eeq
We have changed the normalization, absorbing a factor of 2 into the parameter $\mu$.
The limit $l \rightarrow \infty$ is also not needed in this expression.
For the group element $g$, we also have $g^\dagger \gamma_0 = \gamma_0 \, g^{-1}$.
Using this relation and the representation (\ref{13}, \ref{13a}), we find
\beq
\A = \mu l\, (\Lambda_L)_{A2}\, d X^{A {\dot A}} \, (\Lambda_L)_{{\dot A} 2}
\label{14}
\eeq
where $A, \, {\dot A} = 1, 2$.
We can now define $\pi_{\dot A} = \sqrt{\mu l} \, (\Lambda_L)_{{\dot A} 2}$,
$\omega^A = -i X^{A {\dot A}} \pi_{\dot A}$.
Further, let
\beq
Z = \left( \begin{matrix} \omega^A\\ \pi_{\dot A}\\ \end{matrix} \right),
\hskip .3in
{\bar Z} = Z^\dagger \, \gamma_0 = ({\bar \pi}_A ~~ {\bar \omega}^{\dot A})
\label{16}
\eeq
The action, which is the integral of $\A$, can now be written as
\beq
S = \int \A  = i \int ({\bar \pi}_{A} d \omega^A + {\bar \omega}^{\dot A} d \pi_{\dot A} )
= i \int d\tau\, {\bar Z} \, {\dot Z}
\label{15}
\eeq
From their definition, $Z$, ${\bar Z}$ are seen to obey the condition
${\bar Z} Z = 0$. The strategy now is to regard
all four components of $Z_s$, $s = 1, 2, 3, 4,$ to be independent {\it a priori}
and impose the condition  ${\bar Z} Z = 0$ as a constraint
on the phase space variables for the action (\ref{15}). This
will eliminate the arbitrary parameter $\mu l$ and one of the phases
in $Z$.
Classically, the solution of the constraint
will lead us back to the expression in terms of the group elements.
Quantum theoretically, ${\bar Z}$ is the canonically conjugate variable
and the constraint generates the transformation $Z\rightarrow \lambda \, Z$,
$\lambda \in {\mathbb C} - \{ 0\}$, so that we get a reduction of the phase space
to the
projective twistor space. Particles with nonzero spin can also be described
in the same formalism by relaxing the constraint to some nonzero constant value
for ${\bar Z} Z$.

We now turn to actions for strings which can be viewed as tracing out a two-dimensional surface in spacetime. Since we can regard Minkowski spacetime as the Poincar\'e  group modulo the Lorentz group, we can view strings as maps of a two-dimensional worldsheet into the Poincar\' e group
subject to certain conditions.
The surface area may be regarded as the (wedge) product of two one-forms,
one of them being time-like and the other space-like.
With no additional spin variables, the timelike one-form can be taken as the first term of
$\A$
in (\ref{4}),
\beq
\A =  \biggl[ -i { m l \over 4} \Tr (\gamma_0\, g^{-1} dg ) \biggr]_{l \rightarrow \infty}
= m \, \eta_{\mu\alpha} \Lambda^\mu_{~0} \, dx^\alpha
\label{18}
\eeq
The spacelike one can be chosen to be along any one of the other directions;
we make the choice
\beq
{\cal B} = \biggl[ -i { {\tilde m} l \over 4} \Tr (\gamma_3\, g^{-1} dg ) \biggr]_{l \rightarrow \infty}
= {\tilde m} \, \eta_{\mu\alpha} \Lambda^\mu_{~3} \, dx^\alpha
\label{19}
\eeq
The action, which is the integral of the area element
or the product of these two one-forms is then
\beqar
S &=& \int V_{\alpha\beta}\, dx^\alpha \wedge dx^\beta\nonumber\\
V_{\alpha\beta} &=& M^2 \eta_{\mu\alpha} \eta_{\nu\beta}
( \Lambda^\mu_{~0} \Lambda^\nu_{~3} 
-\Lambda^\nu_{~0} \Lambda^\mu_{~3} )
\label{20}
\eeqar
with $M^2 = {m {\tilde m}/ 2}$. $V_{\alpha\beta}$ obeys the constraints
\beq
V_{\alpha\beta} \, V^{\alpha \beta} = - 2 \, M^4, \hskip .3in
\e^{\mu\nu \alpha\beta}\,V_{\mu\nu} \, V_{\alpha \beta}  = 0
\label{21}
\eeq
As in the case of point-particles, it is possible to treat $V_{\alpha\beta}$
as {\it a priori} independent variables, enforcing the constraints
via Lagrange multipliers.
We can pull back the two-form in (\ref{20}) to the worldsheet 
 to write the action as an integral over the worldsheet
coordinates $\xi^1, \, \xi^2$.
With a Lagrange multiplier for the first of
the constraints in (\ref{21}), this leads to the action
\beq
S = \int V_{\alpha\beta}\, {\del_a x^\alpha \, \del_b x^\beta}\, d\xi^a \wedge d\xi^b
- {1\over 2} \int d^2\xi\sqrt{-g}\, \left[ V_{\alpha\beta} V^{\alpha\beta} + 2\, M^4\right]
\label{22}
\eeq
where $g_{ab}$ is the worldsheet metric. The equation of motion for
 $\sqrt{-g }$ gives the
constraint on $V_{\alpha\beta}$.
Eliminating them by their equations of motion leads to the Nambu-Goto
action
\beq
S = - 2\,M^2 \int d^2\xi\, \sqrt{-\det \rho}
\label{23}
\eeq
where $\rho_{ab} = \eta_{\alpha\beta}  \del_a x^\alpha \,\del_b x^\beta$ is the induced
metric on the worldsheet.

For a null string, we need one lightlike direction and a spacelike direction.
So we make the choice
\beqar
\A &=&  - { i \over 2}\, \mu l\, \Tr \bigl[ (\gamma_0 + \gamma_3 )\, g^{-1} dg  \bigr]
= \mu l\, \eta_{\mu \alpha} (\Lambda^\mu_{~0} + \Lambda^\mu_{~3}) \, dx^\alpha
\nonumber\\
{\cal C} &=&  - {i \over 2} {\tilde m} \sqrt{\mu l }\, \Tr (\gamma_1\, g^{-1} dg )
= {\tilde m} \sqrt{\mu l}\,\eta_{\mu \alpha} \Lambda^\mu_{~1} dx^\alpha
\label{24}
\eeqar
The action is proportional to $\int \A \wedge {\cal C}$ and has the form
\beqar
S&=& M^2 \int V_{\alpha \beta} \, dx^\alpha \wedge dx^\beta
\nonumber\\
V_{\alpha \beta}&=& \eta_{\mu\alpha} \eta_{\nu\beta}
\left[ (\Lambda^\mu_{~0} + \Lambda^\mu_{~3} ) \Lambda^\nu_{~1} 
- (\Lambda^\nu_{~0} + \Lambda^\nu_{~3} ) \Lambda^\mu_{~1} \right]
\label{25}
\eeqar
where $M^2 $ includes factors of ${\tilde m}$ and $\mu l$.
$V_{\alpha\beta}$ now obey the constraints
\beq
V_{\alpha\beta} \, V^{\alpha \beta} = 0, \hskip .3in
\e^{\mu\nu \alpha\beta}\,V_{\mu\nu} \, V_{\alpha \beta}  = 0
\label{26}
\eeq
This is the form of the action obtained in \cite{bal}.

The null string can also be described using twistors, just like the massless particle.
We have already written $\A$ as $i {\bar Z} d Z$.
For ${\cal C}$, using the parametrization (\ref{13a}),
\beqar
{\cal C} &=& -{1\over 2}  {\tilde m} \sqrt{\mu l }\,  \bigl[ (\Lambda_L^*)_{A2}\, dX^{A {\dot A}} \, (\Lambda_L)_{{\dot A} 1}
+ (\Lambda_L^*)_{A1}\, dX^{A {\dot A}} \, (\Lambda_L)_{{\dot A} 2}\bigr]
\nonumber\\
&=& - {i \over 2} ({\bar Z} dW + {\bar W} dZ)
\label{27}
\eeqar
where we have defined another set of twistor variables
\beq
W = {\tilde m} \left( \begin{matrix} -i X^{A{\dot A}} ( \Lambda_L)_{{\dot A} 1}\\ 
(\Lambda_L)_{{\dot A} 1}\\ \end{matrix} \right),
\hskip .3in
{\bar W} = {\tilde m}~ \bigl( (\Lambda_L)^*_{{A} 1} \,\,\, i (\Lambda_L)^*_{{A} 1} X^{A {\dot A}}
\bigr)
\label{27a}
\eeq
The action for the null string may then be taken as
\beq
S = i \int ({\bar Z} dZ ) \wedge ( {\bar Z} dW + {\bar W} dZ )
\label{28}
\eeq
From the definitions, we have the constraints
\beqar
{\bar Z} Z = 0, &\hskip .3in& {\bar W} W = 0\nonumber\\
{\bar Z} W = 0, &\hskip .3in& {\bar W} Z = 0
\label{29}
\eeqar
As before the strategy is to take variables $Z$, $W$ to be independent
{\it a priori}, and then impose the constraints
(\ref{29}) on the phase space corresponding to the action
(\ref{28}). This twistor version of the null string is then identical to
what was obtained in \cite{ilyenko}.

To recapitulate the results of this section briefly: The action for massive and massless
particles , the Nambu-Goto string, the Schild null string are all described in terms
of the Poincar\'e group. The different cases have been separately discussed in the literature
before, but they are brought together here to a single point of view. The purpose is
primarily to set the stage for the rest of this article.

\section{Particle dynamics in AdS$_5$ spacetime}

We now turn to particle dynamics in AdS$_5$ spacetime, which may be considered as 
$SO(4,2)/ SO(4,1)$.
The $SO(4,1)$ generators are given by
\beq
\Sigma_{\mu\nu} = {i \over 4} ( \gamma_\mu \gamma_\nu - \gamma_\nu \gamma_\mu ) ,
\hskip .3in \mu, \nu = 0, 1, 2, 3, 5
\label{30}
\eeq
The remaining generators in the orthogonal complement are $\gamma_\mu$ themselves.
Defining $z = e^{2 \theta}$,  $e^{i \gamma_5 \theta}$ is diagonal
with eigenvalues $\sqrt{z}, \, 1/\sqrt{z}$. A convenient parametrization of a group element
$g \in SO(4,2)$ is
\beq
g = \left( \begin{matrix} \sqrt{z} & i {{\tilde X} /\sqrt{z}}\\
0& {1/ \sqrt{z}}\\ \end{matrix} \right) \, \Lambda
\label{31}
\eeq
where $\Lambda \in SO(4,1)$ is generated by $\Sigma_{\mu\nu}$ in (\ref{30})
and ${\tilde X} = x^0 - {\vec \sigma}\cdot \vx$.
The Cartan-Killing metric on the coset space is given by
\beqar
ds^2 &=& - {R^2 \over 4} \eta^{\mu\nu}\, \Tr( \gamma_\mu \, g^{-1} d g)\,
\Tr( \gamma_\nu \, g^{-1} d g)
\nonumber\\
&=& R^2\, { dx^2 - dz^2 \over z^2}
\label{32}
\eeqar
This is one of the usual forms of the metric on AdS$_5$.
(Here $R$ is a scale parameter.)

For describing massive particles on this spacetime, we can, as usual, use
\beq
\A = -i {m R \over 2} \Tr ( \gamma_0 g^{-1} dg ) = mR\,
\eta_{\mu \nu }\, \Lambda^\mu_{~0} {dx^\nu \over z}
\label{33}
\eeq
where in the second equality we used (\ref{31}). This identifies the momentum as
$p_\nu = \eta_{\mu\nu} \Lambda^\mu_{~0}/z$, which is seen to obey
the mass-shell
condition
\beq
\eta^{\mu\nu} p_\mu p_\nu = {(m R )^2 \over z^2}
\label{34}
\eeq
This agrees with the constraint discussed in \cite{ABT}.  With the inclusion of spin,
the symplectic potential is to be taken as
\beq
\A = -i {m R \over 2} \Tr ( \gamma_0 g^{-1} dg ) + {s_1 \over 2} 
\Tr (\gamma_1 \gamma_2 \, g^{-1} dg ) + {s_2 \over 2} \Tr (\gamma_3 \gamma_5\,
g^{-1} dg ) \label{35}
\eeq
Since the isotropy group for the orbit is $SO(4)$, which is of rank 2,
we need two spin labels
$s_1$ and $s_2$.

We now consider another representation of the Dirac matrices, given by
\beq
\Gamma_0 = \left( \begin{matrix} 1&0\\ 0&-1\\ \end{matrix}\right),
\hskip .2in
\Gamma_i = \left( \begin{matrix} 0&\sigma_i\\ -\sigma_i&0\\ \end{matrix}\right),
\hskip .2in
\Gamma_5 = - i\,\left( \begin{matrix} 0&1\\ 1&0\\ \end{matrix}\right),
\label{36}
\eeq
These are related to the $\gamma$-matrices in (\ref{13}) by a similarity transformation,
$\Gamma = S \gamma S^{-1}$, with
\beq
S = {1\over \sqrt{2}} \left( \begin{matrix} 1 &1\\ 1&-1\\ \end{matrix} \right)
\label{37}
\eeq
Correspondingly, define $S\, g\, S^{-1} = N $. The group element $N$ obeys
$N^\dagger \, \Gamma_0 \, N = \Gamma_0$. With $\Gamma_0$ as given in
(\ref{36}), we see that $N$ can be taken as an element of $SU(2, 2)$.
The symplectic potential for the spinless case becomes
\beqar
\A &=& -i {m R \over 2} \Tr ( \Gamma_0 \, N^{-1} d N)
=  -i {m R \over 2} \Tr \bigl[ (1 +\Gamma_0) \, N^{-1} d N \bigr]\nonumber\\
&=& -i m R \, N^\dagger_{a r} (\Gamma_0)_{rs} \, d N_{s a}\nonumber\\
&=&  -i m R \,\left[  N^\dagger_{a b}  \, d N_{b a} - N^\dagger_{a {\tilde b}} d N_{{\tilde b} a}
\right]
\label{38}
\eeqar
where we have used the fact that $N^{-1} dN$ is traceless.
The indices $a, b$ take values
$1, 2$, while $r, s$ take values 1 to 4, with ${\tilde b} = 1, 2$ corresponding to
$r, s = 3, 4$. Defining
the $2\times 2$ matrices
\beq
\xi_{ba} = \sqrt{2 m R} \, N_{ba} , \hskip .3in \zeta_{ba} = \sqrt{ 2 m R}\, N_{ {\tilde b} a} 
\label{39}
\eeq
we can write
\beq
\A = - {i \over 2} \Tr ( \xi^\dagger d \xi - \zeta^\dagger d \zeta )
\label{40}
\eeq
The constraint on $\xi$, $\zeta$ is given by
$N^\dagger \Gamma_0 N = \Gamma_0$ and translates to
\beq
\xi^\dagger \, \xi - \zeta^\dagger \, \zeta = (2 m R) \, {\mathbb 1}
\label{41}
\eeq
This is a $2 \times 2$ matrix equation. We can regard $\xi$ and $\zeta$ as 
{\it a priori} independent variables with the symplectic potential given by
$\A$ in (\ref{40}) and impose (\ref{41}) as constraints on the phase space.
We now introduce two linear combinations of $\xi$ and $\zeta$ as
\beq
\xi = {{\mathbb U} - i\, {\mathbb W} \over \sqrt{2}}, \hskip .3in
\zeta = {{\mathbb W} - i\, {\mathbb U} \over \sqrt{2}}
\label{42}
\eeq
The action (which is the integral of $\A$) and the constraint become
\beqar
S &=& {1\over 2} \int d\tau\, \Tr ({\mathbb W}^\dagger \, {\dot{\mathbb U}}
- {\mathbb U}^\dagger \, {\dot{\mathbb W}} ) \nonumber\\
{\mathbb U}^\dagger \, {{\mathbb W}} - {\mathbb W}^\dagger \, {{\mathbb U}}
&=& i\, (2 m R) {\mathbb 1} 
\label{43}
\eeqar
We see that we have recovered the twistor description of the massive particle
in AdS$_5$ obtained in \cite{ABT}. (Our ${\mathbb W}$ corresponds to
the ${\tilde {\mathbb W}}$ used in that paper.)

It is also useful to consider the massless particle in AdS$_5$. For this, it is easier to
go back to the first representation of the Dirac matrices and use
the group element $g$. Then, up to an overall
scale factor which will be irrelevant, we can take
\beqar
\A &=& - {i\over 2} \, \Tr \bigl[ ( \gamma_0 + \gamma_3) \, g^{-1} dg \bigr]
= - {i\over 2} \, \Tr \bigl[ ( 1 + \gamma_3 \gamma_0 ) \, {\bar g} dg \bigr]\nonumber\\
&=& -i \, \bigl[ {\bar g}_{2 r} d g_{r 2} + {\bar g}_{3 r} d g_{r3} \bigr]\nonumber\\
&=& -i ({\bar Z} d Z + {\bar W} d W )
\label{44}
\eeqar
where ${\bar g} = g^\dagger \, \gamma_0$ and we have defined
\beqar
Z_r = g_{r 2}, &\hskip .3in& W_r = g_{r 3}\nonumber\\
{\bar Z}_r = {\bar g}_{3 r}, &\hskip .3in& {\bar W}_r = {\bar g}_{3 r}
\label{45}
\eeqar
These obey the constraints
\beqar
{\bar Z} Z = 0, &\hskip .3in& {\bar W} W = 0\nonumber\\
{\bar Z} W = 0, &\hskip .3in& {\bar W} Z = 0
\label{46}
\eeqar
which are identical to (\ref{29}), but, now, for massless particles in
AdS$_5$.

Turning to strings in AdS$_5$, we can use a spacelike one-form of the form
$\Tr (\gamma_3 g^{-1} dg)\sim \eta_{\alpha\beta} \Lambda^\alpha_{~3} dx^\beta/z$ along with
the timelike one-form in (\ref{33}). The action is then of the same form as in
(\ref{22}), with $V_{\alpha\beta}$ as in (\ref{20}), except for the fact that
$\Lambda^\alpha_\mu$ is now an element of $SO(4,1)$ rather than just
$SO(3,1)$. For null strings, one can again use the action (\ref{25}), with
$\Lambda \in SO(4,1)$.

One can also write down a twistor description for null strings. 
Taking the spatial direction to be along
$\gamma_1$, and the null vector as given in (\ref{44}), the action is seen to be of the form
\beq
S = \int ({\bar Z} dZ + {\bar W} dW )\wedge (
 {\bar W} d W' + {\bar W}' dW  -{\bar Z} d Z' - {\bar Z}' dZ)
\label{47}
\eeq
We have introduced two more twistors, corresponding to
$Z' \sim g_{r 1}$, $W' \sim g_{r 4}$. The constraints follow from
${\bar g} \, g = \gamma_0$; explicitly, we have
\beq
{\bar Z}' W = 1, \hskip .3in {\bar Z} W' = 1
\label{48}
\eeq
with all other products of the form (${\overline{\rm twistor}}~\, {\rm twistor}$) vanishing.
(The constraints in (\ref{48}) are complex, the conjugates are obtained as well.)
Given the large number of variables and constraints, this is not an economical way to
describe null strings.

In this section, in terms of the relevant isometry group $SO(4,2) \sim SU(2,2)$,
we have obtained the actions for massive and massless particles and for strings (including the null
string) in AdS$_5$. In particular, the standard co-adjoint orbit action is shown to give,
in a simple straightforward way, the twistor description of massive particles
obtained in \cite{ABT}. We also give the twistor version for massless particles and null strings
in AdS$_5$.

\section{A Chern-Simons approach}

\subsection{Chern-Simons action and co-adjoint orbits}

We now come to the theme of the second
part of this article, namely, relating the co-adjoint orbit actions to the 
Chern-Simons theory and to the CS gravity.
The basic idea is that, generally for any field theory,
the motion of a classical configuration
within the gauge group can be described using a group element
as the collective coordinates. The quantization of the action for these
will generate the dynamics. If we use a Chern-Simons form as the starting
action, then, because of its topological nature, one basically gets trivial dynamics
except for a full group representation. Thus the co-adjoint orbit action should be obtainable
from the CS theory.

To see how this works out in detail,
consider $2+1$ dimensional spacetime of the form
$M \times {\mathbb R}$, where the spatial manifold $M$ has the topology of a disc.
We then consider the action
\beq
S = {k \over 4 \pi} \int \Tr ( A \, dA + {\textstyle{ 2\over 3}} A^3 ) + S_b (A, \psi )
\label{49}
\eeq
Here $A$ belongs to the Lie algebra of some Lie group $G$, $\psi$ is a field
which may be taken as representing degrees of freedom on the boundary of the
manifold. $S_b (A, \psi )$ is the action for $\psi$
on $\del M \times {\mathbb R}$.
The variation of the CS term in (\ref{49}) under a gauge transformation
$A\rightarrow A^g =  g^{-1} A g + g^{-1} dg$ is given by
\beq
{k \over 4 \pi} \int \Tr ( A^g \, dA^g + {\textstyle{ 2\over 3}} A^{g3} )
= {k \over 4 \pi} \int \Tr ( A \, dA + {\textstyle{ 2\over 3}} A^3 )
- \Gamma [g] + {k \over 4 \pi} \oint_b \Tr (A \, dg g^{-1})
\label{50}
\eeq
where
\beq
\Gamma[g] = {k \over 12 \pi} \int \Tr (dg g^{-1} )^3
\label{51}
\eeq
We take $S_b (A, \psi)$ to have a compensating gauge anomaly; i.e.,
we choose $S_b (A, \psi)$ so that
\beq
S_b (A^g, \psi^g) = S_b (A, \psi) - {k \over 4 \pi} \oint_b \Tr (A \, dg g^{-1})
+  \Gamma [g] 
\label{52}
\eeq
Thus the action (\ref{49}) is invariant under all gauge transformations of the fields,
{\it including the transformations $g$ which are not necessarily the identity on the boundary $\del M$}.
The nature of the field $\psi$ and the action $S_b$ are not important for what follows; we may even replace $S_b$ by an effective action obtained by integrating out
$\psi$,
\beq
e^{i S_{\rm eff} (A)} = \int [\D \psi] \, e^{i S_b (A, \psi )}
\label{53}
\eeq
The variation of the action (\ref{49}) is
\beq
\delta S = {k \over 2 \pi} \int \Tr ( \delta A \, F)  + {k \over 4 \pi} \oint \Tr \left(
\delta A \, A \right) + \delta S_b (A, \psi)
\label{54}
\eeq
where $F = dA + A^2$. We see that the equation of motion gives $F = 0$ in the bulk;
we may extend this to the boundary by continuity, with the total current
$(k /4\pi) \e^{ij} A_j + (\delta S_b/\delta A_i) = 0$ as a constraint on the boundary.
Since $F =0 $, $A$ is a pure gauge in the bulk.

Our aim now is to consider singular classical solutions on the disc $M$ of the
form 
\beq
A_i = a_i , \hskip .3in da + a^2 = \sum_{\alpha =1}^N  q_\alpha \,\delta^{(2)}(x - x_\alpha),
\hskip .3in
A_0 = a_0 = 0
\label{55}
\eeq
This corresponds to $N$ magnetic vortices at the points $\vx_\alpha$, with charges $q_\alpha$.
For simplicity, we will take $q_\alpha$ to be in the Cartan subalgebra of $G$, 
so that we may write $da +a^2 = da $. Thus we are considering Abelian magnetic
vortices.
To discuss this case using the Chern-Simons action, we
consider the theory on ${\tilde M}$ which is the disc
$M$ with a number of points
$\vx_\alpha$ removed;
i.e., ${\tilde M} = M - \{ \vx_\alpha\}$.  The boundary of ${\tilde M}$ is 
thus the outer boundary (or $\del M$) at
$\vert\vx\vert \rightarrow \infty$ and a 
set of circles $C_\alpha$, one around each point $\vx_\alpha$.
On ${\tilde M}$, since we have excised small 
neighborhoods around $\vx_\alpha$, we have 
$da = 0$.
The points $\vx_\alpha$ do not lie on the boundary and hence $F= 0$ on the new
boundaries $\{ C_\alpha \}$
as well.
Thus $A$ is a pure gauge, say, $U^{-1} dU$,
and the
solution (\ref{55}) is nonsingular on all of ${\tilde M}$.
Notice that even though $a_i = U^{-1} \del_i U$ on ${\tilde M}$, $U$
is singular at $\vx_\alpha$, since the field strength is nonvanishing at
those points; so they have to be excluded from the manifold
for a nonsingular description.

The general solution of the equations of motion is a pure gauge
on ${\tilde M}$. It may be viewed as the gauge transform of
(\ref{55}) by an element $g \in G$ and can be written as
\beq
A_i = g^{-1} a_i \, g + g^{-1} \del_i g = (U g )^{-1} \del_i (U g) \hskip .3in
A_0 = g^{-1} \del_0 g = (U g )^{-1} \del_0 (U g)
\label{56}
\eeq
$g$ is to be nonsingular everywhere on $M$.
The value of $g$ on the outer boundary
$\del M$ is compensated for by the boundary action
$S_b (A, \psi)$. But the values of $g$ on the new boundaries, i.e.,
at $C_\alpha$ are
edge degrees of freedom which will act as moduli or collective coordinates
for the solution (\ref{55}).
The dynamics of these moduli can be analyzed by evaluating the action on the 
general solutions (\ref{56}). 
The result is thus
\beq
S = S[ a] - {k \over 4 \pi} \sum_\alpha \oint_{C_\alpha} \Bigl[\Tr (a \, dg \, g^{-1})\Bigr]_{\vx_\alpha}
\label{57}
\eeq
Since $a$ has only spatial components, $dg g^{-1}$ in
(\ref{57}) must be $\del_0 g \, g^{-1} dt$.
Now consider shrinking the circles $C_\alpha$ to zero radius.
Since $g$ is nonsingular at $\vx_\alpha$, $ \del_0 g \, g^{-1} dt$ has a limit which we
denote as ${\dot h}^{-1}_\alpha \, h_\alpha dt =
- h^{-1}_\alpha {\dot h}_\alpha \, dt$, where $g(\vx_\alpha, t)=h^{-1}_\alpha \in G$;
$h_\alpha$ can be taken as a time-dependent group element, one element for each vortex.
Further
$\oint_{C_\alpha}  a = q_\alpha$ and the action for the moduli becomes
\beq
S =  {k\over 4\pi} \sum_\alpha \int dt~\Tr (q_\alpha \,  h^{-1}_\alpha\,{\dot h}_\alpha )
\label{58}
\eeq
Writing $q = \sum_a w_a \, t_a$ where $t_a$ are the diagonal generators of the group,
and $w_a$ are suitable weights of a representation,
we see that we have obtained the co-adjoint orbit action.

A couple of remarks are in order at this point.
We used the boundary action $S_b (A, \psi)$ so that we could start 
with a gauge invariant theory on $M \times {\mathbb R}$.
This can be avoided if we take $M$ to be compact without boundary to begin with.
For example, we could consider vortices on $M = S^2$ rather than the disc.
The special solution (\ref{55}) corresponds
to a certain total flux $\sum_\alpha q_\alpha$ 
which is split into the vortices
at the locations $\vx_\alpha$. (If one considers the sphere as embedded in 
${\mathbb R}^3$, this is equivalent to having a monopole of charge
$\sum_\alpha q_\alpha$ at the center.)
After excising the singular points, the boundary
is given by the union of $C_\alpha$; there is no other boundary to consider.
However, since the total flux in nonzero, we will need to consider different
gauge potentials on different coordinate patches with suitable
transition gauge transformations on the overlaps to get a nonsingular 
description. There will also be Dirac quantization conditions on the fluxes.

Also, so far we have only considered introducing collective coordinates for
the dynamics of the vortices within the group. Generally, one could also consider
dynamics or motion of the particles where the locations $\vx_\alpha$ evolve in
time. For the Chern-Simons action, the Hamiltonian is zero and the
worldlines of the vortices can be chosen freely.
(To phrase this another way, one can define a functional integral
for the Chern-Simons action with the insertion of Wilson lines
corresponding to the vortices; the corresponding Hamiltonian
is then precisely what is needed to evolve them along predetermined
worldlines. This was how it was done in \cite{Witten}.)
The only nontrivial dynamics is then the braiding
of the worldlines. 

\subsection{Particles and dynamics in Chern-Simons gravity}

We will now consider using this method for the Poincar\'e or AdS group.
Since the action is a topological one, there is no difficulty in using it for
 noncompact groups such as these.
We will consider two different 
directions in what follows.
First, we simply consider
 the action as in (\ref{49}) and take the gauge group to be
 the Poincar\'e or AdS group of the appropriate dimension.
 This will immediately give the co-adjoint orbit action
 for particles. Thus various actions for particle dynamics considered in the 
 previous sections are obtained for suitable choices of
 the charges in the Cartan subalgebra.  
 
The second, more natural, choice would be to consider
gravity in terms of the Chern-Simons actions.
In any odd spacetime dimension,
one can write a parity-conserving
gravitational action which is the difference of two CS actions.
Generally, this does not lead to Einstein gravity, except in
2+1 dimensions where it does indeed describe
Einstein gravity.
The choice of the group is natural in this approach. Our use of vortex
solutions also ties in
with the well known  observation due to Einstein, Infeld and Hoffmann (EIH),
 where
the field equations for gravity determine the
dynamics of particles, the latter being identified
as singularities of the gravitational field \cite{EIH}.
In their analysis, EIH used
surface conditions around the singularities of the field
to obtain the dynamics,
with the surface at infinity giving the description of the center of mass
motion for the particles. 
More recently, there have been a number of investigations of
the dynamics of a fluid on
the boundary of AdS$_5$ as determined by the bulk field equations \cite{fluid-g}.
Effectively, one considers diffemorphisms of special solutions,
in very much the same way as we consider gauge transformations
of the vortex solutions (\ref{55}).
There have also been some recent studies of vortex-particle duality in 2+1 dimensions.
In some sense, our use of the CS action is the most elementary prototype
of such considerations. In particular, we may think of what follows as 
 the CS gravity analogue of the EIH work. 
Being a CS version of gravity,
there is no real dynamics, so we just end up getting noninteracting particles, 
each described as 
a unitary irreducible representation of the corresponding group.

We will first consider gravity in 2+1 dimensions, for which the relevant group is
$SO(2,1) \times SO(2,1)$. We will refer to these as
$SO(2,1)_L$ and $SO(2,1)_R$ to distinguish them. The generators
of the Lie algebras are denoted by $M_a$ for $SO(2,1)_L$ and $N_a$ for
$SO(2,1)_R$ with the standard commutation rules. A useful
spinorial
matrix
representation is given by
\beqar
&&M_0 = \left[ \begin{matrix} -t_3&0\\ 0&0\\ \end{matrix} \right], \hskip .2in
M_1 = \left[ \begin{matrix} i t_2&0\\ 0&0\\ \end{matrix} \right],\hskip .2in
M_2 = \left[ \begin{matrix} -i t_1&0\\ 0&0\\ \end{matrix} \right]\nonumber\\
&&N_0 = \left[ \begin{matrix} 0&0\\ 0&-t_3\\ \end{matrix} \right], \hskip .2in
N_1 = \left[ \begin{matrix} 0&0\\ 0&-i t_2\\ \end{matrix} \right],\hskip .2in
N_2 = \left[ \begin{matrix} 0&0\\ 0&i t_1\\ \end{matrix} \right]
\label{59}
\eeqar
where $t_a = {\half} \sigma_a$ and $\Tr (M_a M_b ) = \Tr (N_a N_b) = {\half} \eta_{ab}$,
$\eta_{ab} = diag (1, -1, -1)$. Further,
$\Tr (M_a N_b ) =0$.
We also introduce the connections
\beq
A_L = (-i M_a) \, A^a_L = (-i M_a) \left( {\tilde \omega}^a + {e^a\over l} \right),
\hskip .2in
A_R = (-i N_a) \, A^a_R = (-i N_a) \left( {\tilde \omega}^a - {e^a\over l} \right)
\label{60}
\eeq
where we will identify $e^a$ as the frame fields and ${\tilde\omega}^a$ is
related to the spin
connection $\omega^{bc}$ by
\beq
{\tilde \omega}^a = - {1\over 2} \eta^{ak} \e_{kbc}\, \omega^{bc}
\label{61}
\eeq
In (\ref{60}) $l$ is a constant with the dimension of length; it is related to the cosmological
constant.
In terms of the $SO(2,1)$'s,
the generators of Lorentz transformations are
 $L_a = M_a + N_a$ and translations are generated by
$P_a = (M_a - N_a)/l$.
The action for gravity is then given by
\beqar
S &=& -{k \over 4\pi} \left[\int \Tr \left( A dA + {\textstyle{2\over 3}} A^3\right)_L
- \int \Tr \left( A dA + {\textstyle{2\over 3}} A^3\right)_R \right]\nonumber\\
&=&  -{k \over 4\pi \, l} \int d^3x\, \det e~\left[ R - {2 \over l^2} \right]
+ {\rm total~derivative}
\label{62}
\eeqar
where $R$ is the Ricci scalar for the curvature corresponding to $\omega$.
This is the Einstein-Hilbert action, with
Newton's constant of gravity $G$ being related to the level number
of the CS action by $k = (l / 4 G)$.
The standard parity transformation along with
$A_L \leftrightarrow A_R$ leaves the action invariant.

Starting with the first line of (\ref{62}) with the Chern-Simons actions, we can add a
boundary action as in (\ref{49}) and consider
vortices. The resulting particle action will be
as in (\ref{58}), and reads
\beqar
S &=& - {k\over 4\pi} \sum_\alpha \int dt~\left[ \Tr (q_\alpha \,  h^{-1}_\alpha\,{\dot h}_\alpha )_L
- \Tr (q_\alpha \,  h^{-1}_\alpha\,{\dot h}_\alpha )_R\right]\nonumber\\
&=&- {k \over 8 \pi} \int dt\,\sum_\alpha \Bigl[
(q_{L\alpha}  + q_{R\alpha}) \Tr (M_0 - N_0) g_\alpha^{-1} {\dot g}_\alpha ) \nonumber\\
&&\hskip 1in + (q_{L\alpha} - q_{R\alpha}) \Tr (M_0 + N_0) g_\alpha^{-1} {\dot g}_\alpha )\Bigr]
\label{62a}
\eeqar
where
\beq
g = \left( \begin{matrix} h_L&0\\ 0&h_R\\ \end{matrix} \right)
\label{62b}
\eeq
This describes multiparticle dynamics in 2+1 dimensions with
the mass and spin of the particle related to the weights $q_L$, $q_R$ by
\beqar
m &=&  {k \over 8 \pi \, l} (q_L + q_R ) = {q_R + q_L \over 32 \pi G}\nonumber\\
s&=& {k \over 4} (q_L - q_R)
\label{62c}
\eeqar
Here we are considering equal number of vortices for the two groups.
For unequal numbers of vortices, we will get states which are
parity-asymmetric. 

A similar analysis can be done in higher dimensions.
Here we will consider the 5-dimensional case, starting with the
general situation before specializing to CS gravity.
The 5d analogue of the action (\ref{49}) is given by
\beq
S = CS (A) + S_b
\label{63}
\eeq
where $CS(A)$ is the 5d Chern-Simons term,
\beqar
CS(A) &=& - {i k \over 24 \pi^2} \int \Tr \left( A F^2 - {1\over 2} A^3 F + {1\over 10} A^5 \right)
\nonumber\\
&=&- {i k \over 24 \pi^2} \int_{M \times {\mathbb R}} \Tr \left( A \, dA \, dA + {3\over 2}
A^3 \, dA + {3 \over 5} A^5 \right)
\label{ex-2}
\eeqar
and $S_b$ is the boundary action (on $\del M \times {\mathbb R}$)
needed to cancel the anomaly from the CS term.
The bulk equation of motion is given by $F\,F = 0$. This is satisfied by
choosing $A$ as a pure gauge (although this may not be the most general solution).
Since the spatial manifold is four-dimensional, the natural choice for
a singular solution is to take the point-like limit of instantons.
The general solution is thus given by
$A = g^{-1} \, a \, g + g^{-1} dg$
where the potential $a$ is a singular solution (point-like configuration)
with 
nonzero instanton number. The instanton density is concentrated at a set of points
$\{ \vx_\alpha \}$,
so that $A$ is a pure gauge in all of ${\tilde M} \equiv M - \{ \vx_\alpha \} $.
We excise small balls around each singularity to define ${\tilde M}$.
The boundary of the excised ball around $\vx_\alpha$ is $C_\alpha$
which is topologically a three-sphere.
Upon using the general solution for $A$, the action becomes
\beqar
S \!\!&=&\!\!\!CS(a) + S_b (a^g, \psi^g) - {ik \over 240\pi^2} \int \Tr ( dg \, g^{-1})^5\nonumber\\
&&\!\!\!+ {ik \over 48\pi^2} \oint_{\del M} \Tr \left( dg g^{-1} (a da + da a + a^3) + a (dg g^{-1})^3 
- {1\over 2} dg g^{-1} \, a \, dg g^{-1} \, a\right)\nonumber\\
&&\!\!\!- {ik \over 48\pi^2} \sum_\alpha\oint_{C_\alpha} \Tr \left( dg g^{-1} (a da + da a + a^3) + a (dg g^{-1})^3 
- {1\over 2} dg g^{-1} \, a \, dg g^{-1} \, a\right)
\label{ex-4}
\eeqar
The boundary action will cancel the integral around $\del M$ as well as the term
proportional to $\Tr (dg g^{-1})^5$.
Now consider evaluating $CS (a)$. Since 
$a$ is a pure gauge in ${\tilde M}$, we have $da = -a^2$, so that
$CS(a) = -(i k /240\pi^2) \int\Tr (a^5)$. The special solution $a$ is at a fixed time
$t$, with $a_0 = 0$, so $a^5 = 0$ and we get $CS(a) = 0$.
The action for the moduli thus reduces to
\beq
S = - {ik \over 48\pi^2} \sum_\alpha\oint_{C_\alpha} \Tr \left( - dg g^{-1} \, a^3 + a (dg g^{-1})^3 
- {1\over 2} dg g^{-1} \, a \, dg g^{-1} \, a\right)
\label{ex-5}
\eeq
To simplify further, we note that the instanton number is given by
\beq
\nu = - {1\over 8 \pi^2} \int \Tr (F \, F) = - {1\over 8 \pi^2} \oint \Tr ( a da + {2\over 3} a^3)
= {1\over 24 \pi^2} \oint \Tr (a ^3)
\label{ex-6}
\eeq
where we have used the fact that $a$ is a pure gauge on ${\tilde M}$ including
on $C_\alpha$.
We will consider point-like instantons in an $SU(2)$ subgroup of the gauge group
$G$. Further we consider the standard embedding of $SU(2)$ in $G$ and
take
$a = U^{-1} dU $, with
with
\beq
U = \phi^0 + i \sigma_i \, \phi^i, \hskip .3in (\phi^0)^2 + (\phi^i \phi^i ) = 1
\label{ex-7}
\eeq
In this case, we can work out $a^3$ as
\beq
a^3 = t_1~\e_{\mu\nu\alpha\beta} ~\phi^\mu d\phi^\nu d\phi^\alpha d\phi^\beta
, \hskip .2in
t_1 =  \left[ \begin{matrix} {\mathbb 1} &0 \\
0& 0\\
\end{matrix} \right]
\label{ex-15}
\eeq
The winding number for the $S^3 \rightarrow S^3$ map,
corresponding in our case to the map $U: C_\alpha \rightarrow SU(2)$, is
given by
\beq
Q[u] = {1\over 12 \pi^2} \int  \e_{\mu\nu\alpha\beta} ~\phi^\mu d\phi^\nu d\phi^\alpha d\phi^\beta
\label{ex-15a}
\eeq

We can now start simplifying the terms in (\ref{ex-5}). 
Using (\ref{ex-15}) , (\ref{ex-15a}), the first term of (\ref{ex-5}) can now be written as
\beqar
  {ik \over 48\pi^2} \sum_\alpha\oint_{C_\alpha} \Tr  (dg g^{-1} \, a^3)
  &=&   {ik \over 48\pi^2} \sum_\alpha\int dt \oint_{C_\alpha}  \Tr  (\del_0 g g^{-1} \, a^3)\nonumber\\
  &=&{i k \over 4} \sum_\alpha\int dt \Tr  (t_1 \,\del_0 g g^{-1}) Q_\alpha
  \label{ex-16}
\eeqar
(A similar result is obtained for other $SU(2)$ subgroups of $G$ as well.)
Now consider the next term in (\ref{ex-5}), namely,
$\oint \Tr [a \, (dg g^{-1})^3 ]$. Separating out the time-derivative
part, we are considering terms of the form
\[
\oint \Tr [a \, (dg g^{-1})^2- dg g^{-1} a dg g^{-1} + (dg g^{-1})^2 a] \,\del_0 g g^{-1} dt
\]
where the $d$'s denote differentiation with respect to the angular coordinates
of the three-sphere surrounding $\vx_\alpha$. These derivatives must go to zero
as we shrink the spheres to zero radius to ensure that $g$ is nonsingular.
Thus the contribution of this term is zero. A similar argument can be made for the 
last term in (\ref{ex-5}).
The action for the moduli is thus
\beq
S =  {i k \over 2} \sum_\alpha \oint_{C_\alpha} \left[ Q_\alpha \,\Tr (t_1 \del_0 g_\alpha 
\, g_\alpha^{-1} )\right]
\label{ex-17}
\eeq

We now apply this line of reasoning
to the case of $SO(4,2)\sim SU(2,2)$ which is appropriate for CS gravity
with a cosmological constant
in 5 dimensions. 
The relation between the gauge fields and the frame fields and spin connection is
given by
\beq
A_L = -i \left[ {\half} \omega^{ab} \Sigma_{ab} + {\half } e^a \Gamma_a\right],
\hskip .3in
A_R = -i \left[ {\half} \omega^{ab} \Sigma_{ab} - {\half} e^a \Gamma_a\right]
\label{ex-18}
\eeq
where $\Sigma_{ab} $ generate the $SO(4,1)$ group and $\Gamma_a$ correspond
 to the coset directions. The explicit form of these matrices are as given in
 (\ref{13}, \ref{30}, \ref{36}).
 The action is given by
 \beq
 S = CS(A_L) + S_{bL} - \{ CS(A_R) + S_{bR}\}
 \label{ex-18a}
 \eeq
Consider first $CS(A_L)$. 
There are two $SU(2)$ sugbroups generated by
$i ({\half} \e_{ijk} \Gamma_i \Gamma_j \mp \Gamma_5 \Gamma_k )/4$, corresponding to
 the upper and lower $2\times 2$ block diagonal matrices in the chosen representation.
We consider point-like
instantons in these two $SU(2)$ subgroups.
 We can then adapt (\ref{ex-16}) to the present case; the action for the moduli is then
 \beqar
 S_L &=& {i k \over 4} \sum_\alpha\int dt\,\Tr  \left( {1+ \Gamma_0 \over 2} \,
 \del_0 g g^{-1}\, Q^{(1)}_\alpha + {1- \Gamma_0 \over 2} \,
 \del_0 g g^{-1}\, Q^{(2)}_\alpha \right)\nonumber\\
 &=&- {i k \over 8} \sum_\alpha (Q_\alpha^{(1)}- Q_\alpha^{(2)} ) \int dt\, \,
 \Tr \,(\Gamma_0\, h_\alpha^{-1} {\dot h}_\alpha )
 \label{ex-20}
 \eeqar
 We have made the replacement $g \rightarrow h^{-1}$ to agree with the notation
 of (\ref{58}) and (\ref{61}).
 We can parametrize the group element as in (\ref{31}), namely,
 \beq
h_L = S^{-1}\,g = S^{-1}\left( \begin{matrix} \sqrt{z} & i {{\tilde X} /\sqrt{z}}\\
0& {1/ \sqrt{z}}\\ \end{matrix} \right) \, \Lambda
\label{ex-21}
\eeq
where $S$ is the matrix relating $\Gamma$'s to $\gamma$'s.
The action now reduces to 
\beq
S_L = \sum_\alpha \int {k \over 4}  (Q_\alpha^{(1)}- Q_\alpha^{(2)} )\,
\eta_{\mu\nu} \Lambda^\mu_{~0} \, {dx^\nu \over z}
\label{ex-22}
\eeq
A similar result holds for the $CS(A_R)$ part. In this case, we have 
 \beq
h_R = S^{-1}\,g' = S^{-1}\left( \begin{matrix}  {1/ \sqrt{z}} &0\\
- i X/\sqrt{z}& \sqrt{z}\\ \end{matrix} \right) \, \Lambda, \hskip .3in
X = x^0 + \sigma\cdot \vx
\label{ex-23}
\eeq
The action for this part is then
\beq
S_R = - \sum_\alpha \int {k \over 4}  (Q_\alpha^{(1)}- Q_\alpha^{(2)} )\,
\eta_{\mu\nu} \Lambda^\mu_{~0} \, {dx^\nu \over z}
\label{ex-24}
\eeq
The total action, given by $S_L - S_R$ is thus of the form (\ref{33}), as is appropriate for
particles in AdS$_5$, with the identification
\beq
m R = {k \over 2}  (Q_\alpha^{(1)}- Q_\alpha^{(2)} )
\label{ex-25}
\eeq
This result does not include spin. For spin, we will need to
consider multi-instantons taking account of the placement of the $SU(2)$'s
within the larger group. A simple way to see this is to consider merging
two of the points $\vx_\alpha$, say $\vx_1 $ and $\vx_2$, into a single one. This is equivalent to
setting $h_1 = h_2$. However, there are different placements of the $SU(2)$ subgroups 
in $SU(2,2)$, related by Weyl group transformations. In particular, we have
\beq
T_1^{-1} \, \Gamma_0 \, T_1 = i \Gamma_1 \Gamma_2, \hskip .3in
T_2^{-1} \, \Gamma_0 \, T_2 = i \Gamma_3 \Gamma_5
\label{84}
\eeq
where
\beq
T_1 = \left[ \begin{matrix}
1&0&0&0\\ 0&0&1&0\\ 0&-1&0&0\\ 0&0&0&1\\ \end{matrix} \right], \hskip .3in
T_2 = \left[ \begin{matrix}
1&0&0&0\\ 0&0&0&1\\ 0&-1&0 &0\\ 0&0&-1&0\\ \end{matrix} \right]
\label{85}
\eeq
Thus if we consider merging the two points with the identification $h_1 = h_2$
and $a_{(2)} = T_1^{-1} \, a_{(1)}\, T_1$, we get the action
\beq
S = - {i k \over 8}\int dt\,  \Bigl[ (Q_1^{(1)}- Q_1^{(2)} )  \,
 \Tr \,(\Gamma_0\, h_1^{-1} {\dot h}_1) + 
  (Q_2^{(1)}- Q_2^{(2)} )  \,
 \Tr \,(i \Gamma_1 \Gamma_2 \, h_1^{-1} {\dot h}_1)\Bigr] + \cdots
 \label{86}
 \eeq
By considering a number of such mergers, with different values of
$Q^{(1)}$ and $Q^{(2)}$, we can get a general combination of
$ \Tr \,(\Gamma_0\, h_1^{-1} {\dot h}_1) $, 
$ \Tr \,(i \Gamma_1 \Gamma_2\, h_1^{-1} {\dot h}_1) $ 
and $ \Tr \,(i \Gamma_3 \Gamma_5\, h_1^{-1} {\dot h}_1) $, thus giving a
general action
for particle dynamics.
While this does show the role of the choice of the $SU(2)$ subgroups,
there should be a more satisfactory way of obtaining the general
co-adjoint orbit action, following directly from the multi-instanton solutions.
This is currently being explored and will be deferred to a future publication.

To summarize the results of this section, we considered the dynamics which
follows from the use of singular solutions of the CS theory,
where a gauge transformation furnishes the moduli for the singularities.
This is very similar to the strategy of Einstein, Infeld and Hoffmann,
who derived multiparticle dynamics from the general theory of relativity,
treating particles as singularities of the gravitational field.
We considered the 3d CS theory. We also applied the same strategy to
CS gravity in 3 and 5 dimensions. Clearly there are still many points to be clarified and
elaborated on such as the use of more general solutions including various types of gravitational instantons as well as solutions which can lead to interacting particles in five dimensions,
issues related to the orientations
of various subgroups,
as well as the inclusion of nontopological terms, i.e., going
beyond the CS action. The connection to fluid dynamics, alluded to earlier, is
another interesting avenue to explore. These issues will be taken up in future.

\section*{Acknowledgements}

VPN thanks A.P. Balachandran for discussions.
This research was supported in part by the U.S.\ National Science
Foundation grant PHY-1519449
and by PSC-CUNY awards.


\begin{thebibliography}{99}

\bibitem{geom} Geometric quantization and the co-adjoint orbit actions
have a long history. See for example,
J.-M. Souriau, \CMP~{\bf 1}, 374 (1966);
{\it Structure des Syst\`emes Dynamiques} (Dunod, Paris, 1970);
B. Kostant, Quantization and Unitary Representations, in 
{\it Lectures in Modern Analysis and Applications III}, C.T. Taam (ed.), 
Lecture Notes in Mathematics {\bf 170}, pp. 87-208 (Springer, Berlin, 1970);
A.A. Kirillov, {Geometric Quantization} in {\it Dynamical Systems IV},
V.I. Arnold and S.P. Novikov (eds.),
Encyclopedia in Mathematical Sciences, Volume 4, pp. 139-176,
(Springer, Berlin, 2001).

\noindent More recent general references on geometric quantization include:\\
N.M.J. Woodhouse, {\it Geometric Quantization}, Clarendon Press (1992); 
J. Sniatycki, {\it Geometric Quantization and Quantum Mechanics}, Springer-Verlag (1980);
S.T. Ali and M. Englis,
{\it Quantization Methods: A Guide for Physicists and Analysts}, arXiv:math-ph/0405065;
M. Blau, {\it Symplectic Geometry and geometric quantization},
{\verb+http://www.blau.itp.unibe.ch/Lecturenotes.html+ }; B.C. Hall, {\it  Quantum Theory for Mathematicians}, Springer (2013);
P. Woit, {\it Quantum Theory, Groups and Representations: An Introduction},
Springer (to be published), available at\\
\verb+http://www.math.columbia.edu/%7Ewoit/QM/qmbook.pdf+;
V.P. Nair, {\it Elements of Geometric Quantization
\& Applications to Fields and Fluids}, Lectures at the Second Autumn School on
High Energy Physics and Quantum Field Theory, Yerevan, Armenia,
October 2014, arXiv:1606.06407.


\bibitem{ABT} A.S. Arvanitakis, A.E. Barns-Graham and P.K. Townsend, \PRL~
{\bf 118}, 141601 (2017).

\bibitem{zanelli} For a review of CS gravity, see
J. Zanelli,
{\it Lecture notes on Chern-Simons (super-)gravities. Second edition (February 2008)},
arXiv:hep-th/0502193.

\bibitem{EIH} A. Einstein, L. Infeld and B. Hoffmann, Ann. Math. {\bf 39}, 65 (1938);
A. Einstein and L. Infeld, Can. J. Math. {\bf 1}, 209 (1949).

\bibitem{fluid-g}
S. Bhattacharyya, V. E. Hubeny, S. Minwalla, and M. Rangamani, 
JHEP 0802, 045 (2008), arXiv:0712.2456 [hep-th]; for a review, see 
M. Rangamani, {\it Gravity and hydrodynamics: Lectures on
the fluid-gravity correspondence}, \CQG~{\bf 26}, 224003 (2009).

\bibitem{fluid-us} B. Bistrovic,
R. Jackiw, H. Li, V.P. Nair and S.Y. Pi, \PR~{\bf D67}, 025013 (2003);
V.P. Nair, R. Ray and S. Roy, 
\PR~{\bf D86}, 025012(2012); for a review, see
R. Jackiw, V.P. Nair, S-Y. Pi and
A.P. Polychronakos, J. Phys. A: Math. Gen. {\bf 37}, R327 (2004).

\bibitem{KN} D. Karabali and V.P. Nair, \PR~{\bf D90}, 105018 (2014).

\bibitem{bal} A.P. Balachandran, F. Lizzi and G. Sparano,
\NP~{\bf B263}, 608 (1986);
\NP~{\bf B277}, 359 (1986);
see also, A.P. Balachandran, G. Marmo, B.-S. Skagerstam and A. Stern, {\it Gauge symmetries and fibre
bundles: applications to particle dynamics}, Lecture Notes in Physics, vol. 188 (Springer, 1983).

\bibitem{ilyenko} T. Shirafuji, Prog. Theor. Phys. {\bf 70}, 18 (1983);
K.Ilyenko, \NP~{B (Proc. Suppl.)} {\bf 102-103}, 83 (2001);
L. Mezincescu, A.J. Routh and P.K. Townsend, J. Phys. A: Math. Theor. {\bf 49},
025401 (2016).

\bibitem{Witten} E. Witten, \CMP~{\bf 121}, 351 (1989).


\end{thebibliography}
\end{document}